\documentstyle[version2,preprint,aps]{revtex}  
\begin{document}
\draft
\begin{title}
 Dissipation of the $^3$He $A\to B$ Transition
\end{title}
\author{Peter Kost\"adt and Mario Liu}
\begin{instit}
 Institut f\"ur Theoretische Physik, Universit\"at Hannover,
 3000 Hannover 1, Germany
\end{instit}
\begin{abstract}
 A rigorous hydrodynamic theory of the $A$-$B$ transition is presented.
 All dissipative processes are considered. At low interface velocities,
 those occurring on hydrodynamic length scales, not considered hitherto,
 are most probably the dominant ones.
\end{abstract}
\pacs{67.50.Fi, 68.10.-m}

\narrowtext
The $A\to B$ transition of superfluid $^3$He is rather remarkable. If
undercooled sufficiently, it takes place with a spectacularly fast rate,
and is accompanied by magnetic signals that can only be called bizarre
\cite{boyd92}. However, no-one was left wondering about the damping
mechanism, as Yip and Leggett \cite{yip86} instantly identified it:
The superfluid order-parameter varies rapidly within the interface,
transforming one phase into the other. This scatters quasiparticles
(Andreev scattering) and constitutes a restoring force. The balance between
this and the driving force $\Delta\mu$ (the difference in chemical potential
of the two phases) yields a terminal interface velocity $\dot u$ that can be
compared with experimental data \cite{boyd92}. Further and more detailed
microscopic calculations \cite{legg92} confirmed Andreev scattering as the
source of damping; also, the magnetic signals were recently deciphered
\cite{paliu92}.

All this, one should think, holds for the hypercooled regime, with an
undercooling $\varepsilon\equiv 1-T/T_{AB}\agt 0.5\,\%$. With $\varepsilon$
smaller, the latent heat would warm up the $B$-phase, and
render it thermodynamically unstable again. So $\dot u$ is much slower and
limited (instead of by Andreev scattering) by how efficiently the latent
heat can be removed from the interface region. (A difficult,
non-local problem notorious from more mundane interfaces such as snow
flakes.) This is quite wrong: In superfluid $^3$He,
there is neither a transition regime limited by heat transfer, nor a sudden
onset of hypercooled phase-transition. Rather, it is the second-sound
velocity $c_2$ that separates two different types of transitional behavior.
For $\dot u\ll c_2$, or $\varepsilon\alt 2\,\%$, second sound is very
efficient in removing the latent heat, which therefore can not be the
limiting factor. What is more, phase coherence across the interface
equalizes the chemical potential, eliminating $\Delta\mu$ as the driving
force. A hydrodynamic consideration \cite{graliu90} shows instead an
interface driven by $\Delta T$ and damped by the Kapitza resistance.
Curiously, the growing $B$-phase is in this regime of `phase-coherent
transition' colder than the receding $A$-phase. When $\dot u$ greatly exceeds
$c_2$, starting at $\varepsilon\approx$~20-30\,\%, second sound is in
comparison too slow to transfer appreciable amount of heat.
Only then does the original scenario of hypercooling re-emerges.

Following all the microscopic theories \cite{yip86,legg92}, the
hydrodynamic consideration \cite{graliu90} also contains the starting
assumption that the dissipation accompanying the $A$-$B$ transition occurs
within the mean free path $\xi_f$ of the interface, and that no dissipative
temperature variations exist on hydrodynamic length scales. This universal
assumption is most probably incorrect.
To understand why, we first examine the case of a
stationary interface between superfluid $^3$He and a vessel wall,
through which heat but no mass is transferred. Generally, the effective,
measured resistance here is the sum of two contributions \cite{graliu87},
$\kappa_e^{-1}=\kappa^{-1}+\kappa_{sq}^{-1}$. The first accounts for the
microscopically fast drop $\Delta T$ across the interface, the second stems
from the `sq-mode', a hydrodynamically slow variation
$\delta T\exp(-|x|/\lambda_{sq})$ in the superfluid. Due to the enormous
extent of the decay length ($\lambda_{sq}\sim T_f/T_c$ is at least 250 times
the mean free path, usually much larger) the effective resistance
$\kappa_e^{-1}$ is dominated by $\kappa_{sq}^{-1}$ \cite{sun89}.
Consequently, $\delta T\gg\Delta T$. Going back to the moving $A$-$B$
interface, it is clear that something akin the sq-mode could also exist there.
As we shall see, this is indeed the case. And since this (what we continue
to call) sq-mode has, for $\dot u\ll c_2$, essentially the same spatial
extent, it is here probably also the dominant source of dissipation.

What is more, there is some indication that, independently, $\Delta T\to 0$.
Recently, Schopohl and Waxman \cite{scho92} considered a moving interface,
between the $A$ and $B$-phase that are in equilibrium otherwise. In contrast
to all previous microscopic calculations \cite{yip86,legg92} that are
perturbative in essence, they have obtained an exact solution, in the
ballistic limit, with an essential singularity at $\dot u=0$. Amazingly,
they found this motion to be (up to a fairly high critical velocity)
little damped \cite{nine}. As will be shown below, the immediate consequence
of this is a diverging Kapitza conductance, and $\Delta T\to 0$
for $\dot u\ll c_2$. In other words, if this finding can be verified,
Andreev scattering as a dissipative source is eliminated altogether, while
the hydrodynamic variation of temperature and counterflow becomes the only
mechanism to prevent the transition rate $\dot u$ from diverging.

In this paper, we present the general hydrodynamic theory of the $A$-$B$
transition. All dissipative mechanisms that may occur are considered.
Despite a rather different language, they include collisions and scattering
of quasiparticles, both among themselves and at the interface.
More specifically, we derive the general boundary conditions connecting two
strongly coupled superfluids and calculate the temperature and counterflow
fields. Although the hydrodynamic
theory is never complete by itself, our results do provide a rigorous
framework for the more detailed, and rather more complicated, microscopic
theory. In fact, the latter is essentially reduced to the calculation of
three Onsager coefficients.

Concrete results are obtained for the two limits $\dot u\ll c_2$ and
$\dot u\gg c_2$. In the first case of slow, phase-coherent transition, the
general temperature variation contains two exponential decays
$\delta T_{sq}^{A,B}\exp(-|x|/\lambda_{sq})$ in the respective phase, and a
discontinuity $\Delta T$ at the interface ($x=0$); cf Fig.~1.
While $\delta T_{sq}^{A,B}$ stem predominantly from collisions among
quasiparticles, $\Delta T$ accounts for their scattering at the interface.
(The counterflow is not independent, $\delta w^{A,B}\sim\delta T^{A,B}$.)
The decay length $\lambda_{sq}$ is a function of known bulk coefficients;
to lowest order in $\dot u/c_2$ it is equal to the decay length, mentioned
above, of $^3$He close to a vessel wall, and hence large.
The interface motion is damped by a total, effective Kapitza resistance, which
is a series of three consecutive resistive elements, each causing one of the
temperature drops.
The amplitudes of these are determined by three Onsager coefficients, unknown
in size. So it is these three numbers that need to be calculated, or measured.
Until now, it was assumed that $\delta T_{sq}^{A,B}=0$, leaving $\Delta T$ to
account for the total dissipation. If, conversely, $\Delta T$ is negligible as
mentioned, one may (for lack of better knowledge) assume
$\delta T_{sq}^A\approx\delta T_{sq}^B$. Then the total Kapitza resistance
depends only on one parameter, which can be determined from the
experimental data on $\dot u$, as we shall do.

If $\dot u\gg c_2$, the varying fields of the temperature and counterflow
$\sim\delta T_d^A$, $\delta w_d^A$ are independent, diffusive and decay only
into the $A$-phase; cf Fig.~2. The decay length is smaller by the
factor $c_2/\dot u$. Neither the temperature nor the chemical potential is
continuous across the interface; $\Delta T$, $\Delta\mu\not= 0$. There is no
special reason why $\Delta T$ should be much larger or smaller than
$\delta T_d^A$. The interface motion is damped by a total growth coefficient
which, however, contains additive as well as multiplicative contributions.

It is noteworthy that all the results of Ref.~\cite{graliu90} remain
asymptotically valid (ie for distances from the interface that are large
compared to all decay lengths), if one substitutes the respective resistance
with the total Kapitza and growth resistance obtained here. As will be
explained in details below, this is connected to the fact that one can
consider an effective interface, hydrodynamically wide, that includes all
the temperature and counterflow variations; cf the dotted lines of
Figs.~1. and 2. Then, of course, the original assumption that
dissipation takes place only within the interface is again correct.
In this work, for lack of space, we do not consider the effects of lateral
walls, which lead to an $R$-dependence of the terminal velocity $\dot u$,
as observed \cite{boyd92}.

An interface in motion can be viewed as condensate and quasiparticles
traversing the interface. It is plausible that the condensate should not
be damped. But the Schopohl-Waxman solution \cite{scho92} shows that even
the quasiparticles are little damped in equilibrium, despite considerable
Andreev scattering. This is a surprising result, and as the following
arguments show, has direct bearing on the non-equilibrium properties of the
interface: Usually, the temperature establishes itself on the scale of the
mean free path $\xi_f$, and the temperature gradient $\nabla T$ has a
hydrodynamic scale much larger.
However, across a strongly resistive obstacle of microscopic dimension
$\xi\ll\xi_f$, the change in the temperature will be on the same scale
$\xi$ and can be hydrodynamically accounted for as a discontinuity
$\Delta T$. The $A$-$B$ interface, with a width of order correlation
length $\xi_c\ll\xi_f$, was taken as just such a microscopic obstacle
\cite{yip86,legg92}. And its resistivity (outside a very narrow range
next to the normal-superfluid transition) would come mainly
from Andreev scattering of ballistic quasiparticles. If this is indeed
inoperative in equilibrium, it cannot turn into a strongly resistive
mechanism ever so slightly off equilibrium. The temperature gradient
will therefore have normal, hydrodynamic values, and
$\Delta T\approx\xi_f\nabla T$ vanishes. A more formal line of arguments
that shall be published elsewhere leads to the same conclusion. Further away
from equilibrium, when $\dot u$ becomes comparable to, or much larger than,
the second sound velocity $c_2$, $\Delta\mu$ builds up across the
interface \cite{graliu90}. This would constitute the microscopic
obstacle lacking at $\dot u\ll c_2$, and an accompanying $\Delta T$
can no longer be ruled out by the same argument.

We start our hydrodynamic consideration with the general solution that is
stationary in the rest frame of the interface. For both $\dot u\ll c_2$ and
$\dot u\gg c_2$, we may linearize the hydrodynamic equations \cite{vw} as
in Ref.~\cite{graliu90}, with respect to the variables (i)
$w\equiv\rho_s(v_n-v_s)/\rho_n\equiv v_n-g/\rho$ and (ii) $T^{A,B}-T_i$,
the deviation of the temperature in the respective phase from the
initial temperature $T_i$. Retaining terms of first order in $\dot u/c_2$,
the solution (in both phases) for $\dot u\ll c_2$ is
\FL
\begin{mathletters}\begin{eqnarray}
  T^{A,B}&=&T_i+\delta T_2^{A,B}+\delta T_{sq}^{A,B}
            \exp(\mp x/\lambda_{sq})\;, \label{sol1a}           \\
  w^{A,B}&=&\pm\frac{c_2\sigma_T}{\sigma}\left(\delta T_2^{A,B}
            -\sqrt{\frac{\lambda_T}{\lambda_w}}
            \delta T_{sq}^{A,B}\exp\Bigl(\mp\frac{x}{\lambda_{sq}}\Bigr)
            \right) \label{sol1}
\end{eqnarray}\end{mathletters}

Notations and explanations: Upper sign refers to the $A$-phase, here and
below. $\delta T_2^{A,B}$: amplitude of the second sound step-function
in the respective phase \cite{graliu90}. Although the steps are at
$\pm c_2 t$, $t\to\infty$ must be set, since Eqs.~(1) display
the stationary solution. (Here and below, if the context is clear, the
superscripts $A$ and $B$, eg in $c_2^A$, will be suppressed.)
$\delta T_{sq}^{A,B}$: amplitude of the sq-mode, source of hydrodynamic
dissipation and resistance. $\sigma$: entropy per unit mass,
$\sigma_T\equiv\partial\sigma/\partial T$.
$\lambda_{sq}^{A,B}=2(\lambda_T\lambda_w)^{1/2}\mp(\lambda_T+\lambda_w)
\dot u/c_2$: the sq-decay length for a moving interface,
$\lambda_T\equiv k/(2c_2\rho T\sigma_T)$,
$\lambda_w\equiv [(4/3)\eta-\rho(\zeta_1+\zeta_4)+\zeta_2+\rho^2\zeta_3]
            \rho_s/(2\rho\rho_n c_2)$,
where the heat conductance $k$ and the viscosities $\eta$, $\zeta_{1-4}$
are defined in the usual way \cite{vw}, neglecting the anisotropy.

The solution for $\dot u\gg c_2$ is, to lowest order in $c_2/\dot u$
\FL
\begin{mathletters}\begin{eqnarray}
  T^B&=&T_i+\delta T^B\,,\;
        T^A=T_i+\delta T_d^A\exp(-\dot u x/2c_2\lambda_T)\;, \label{sol2a}\\
  w^B&=&\delta w^B\,,\; w^A=\delta w_d^A\exp(-\dot u x/2c_2\lambda_w)\;.
  \label{sol2}
\end{eqnarray}\end{mathletters}

$\delta T_d^A$ and $\delta w_d^A$ are respectively the diffusive modes
of a moving interface \cite{brenn}. Next order terms in $c_2/\dot u$ mix
these two modes.

Each of the four amplitudes of Eqs.~(1,2) are to be
determined in conjunction with $\dot u$ from boundary conditions, better:
connecting conditions  (CoCos). The general structure of the CoCos depends,
as do bulk hydrodynamic theories, only on the conserved quantities and the
spontaneously broken symmetries on both sides of the interface
\cite{graliu87}. In our case, the CoCos are given by the continuity of the
fluxes for energy, mass and momentum, the phase coherence across the
$A$-$B$ interface, and the surface entropy production rate $R_s$.
These are respectively
\FL
\begin{mathletters}\begin{eqnarray}
  \Delta Q&=&\Delta g=\Delta (p+\pi+\pi^D)=0\;,   \\
  \Delta\dot\varphi&\equiv&-\Delta(\mu+v_n v_s+z^D)=0\;, \\
  R_s=\langle f\rangle\Delta T&+&g\Delta(\mu+z^D)+
        \Delta\bigl(v_n(\pi+\pi^D-\rho z^D)\bigr)    \label{rs}
\end{eqnarray}\end{mathletters}

($p$: pressure; $\pi$: the nonlinear part of the stress tensor,
$\pi^D$ its dissipative part; $z^D$: dissipative part of the Josephson
equation.) Eqs.~(3) reduce to the expressions of Ref.~\cite{graliu90} if one
excludes dissipative terms (with superscript $D$). All quantities are
defined in the interface system;
$\langle\ \rangle$ and $\Delta$ denote average and difference across the
interface; and all suppressed indices point along the interface normal.
Neglecting $\Delta\rho/\rho\sim 10^{-8}$ and for time scales slow
compared to first sound velocity, $g=-\rho\dot u$ holds and $\Delta g=0$
is always satisfied. Linearizing the other CoCos, for the weakly
supercooled case $\dot u\ll c_2$, with respect to $w$, $\dot u$ and
$\Delta T$, we obtain
\begin{mathletters}\begin{eqnarray}
 \Delta f&=&0\;,\ \Delta(p+\pi^D)=0\;,\ \Delta(\mu+z^D)=0\;, \label{cc1a} \\
 \langle f\rangle&=&\kappa\Delta T\;,
 \ v_n^{A,B}=\mp\alpha_{A,B}(\pi^D-\rho z^D)^{A,B}\;.   \label{cc1}
\end{eqnarray}\end{mathletters}

Eqs.~(\ref{cc1}) are the Onsager relations that follow from $R_s$ of
Eq.~(\ref{rs}). The last two CoCos are new: Neglecting dissipative terms, they
would vanish (first in $R_s$ and hence altogether). Positivity of
entropy production requires $\alpha_{A,B}>0$; the cross terms, such as
$v_n^A\Delta T$ in $R_s$, are neglected for simplicity. The values of
$\alpha_{A,B}$ determine the rate of dissipation both within the interface
(contribution to $R_s$) and outside (contribution from the sq-mode).
The latter with a vastly larger width $\sim\lambda_{sq}$, dominates.

We expand Eqs.~(4) around $T_i$ and denote all thermodynamic quantities
at that temperature. To distinguish, a square bracket with index $i$ is
added, eg $[\Delta\mu]_i\equiv\mu_B(T_i,\,p_i)-\mu_A(T_i,\,p_i)$; while
in Eqs.~(\ref{cc1a}) $\Delta\mu\equiv\mu_B(T_B,\,p_B)-\mu_A(T_A,\,p_A)$. With
$\tilde\alpha\equiv\alpha c_2\rho\rho_n/\rho_s+(\lambda_T/\lambda_w)^{1/2}$,
the results are
\FL
\begin{mathletters}\begin{eqnarray}
 & &\delta T_2^{A,B}={\textstyle\frac{1}{2}}\Bigl[\mp\Delta\mu/
                     \langle\sigma\rangle-(\dot u/c_2)\Delta\sigma/
                     \langle\sigma_T\rangle\Bigr]_i, \\
 & &\delta T_{sq}^{A,B}=\mp\tilde\alpha_{A,B}^{-1}\Bigl[{\textstyle\frac{1}{2}}
                       \Delta\mu/\langle\sigma\rangle+(\dot u/c_2)
                        \langle\sigma\rangle/\langle\sigma_T\rangle\Bigr]_i,
                        \label{res1}   \\
 \dot u&=&\left(\frac{-2\kappa}{(\tilde\alpha_A^{-1}+\tilde\alpha_B^{-1})
               \kappa+\rho c_2\langle\sigma_T\rangle_i}-1\right)
               \left[\frac{c_2\langle\sigma_T\rangle
                       \Delta\mu}{2\langle\sigma\rangle^2}\right]_i.
\end{eqnarray}\end{mathletters}

The extended part $\delta T_2^{A,B}$ of the temperature field agrees with
that of Ref.~\cite{graliu90}, in which the dissipative terms were neglected.
To understand why this is not an accident and what the
essence of the new information here is, we need
to address the concept of the effective CoCo. Since the CoCos are, as
emphasized, quite generally valid, we have a certain discretion towards
the choice of the interface width: It can be either microscopic, of order
$\xi_f$, or it can be hydrodynamic, somewhat
larger than $\lambda_{sq}$. Eqs.~(4), such as they stand, are
the proper CoCos for the microscopic interface, it provides complete
information on $\dot u$, the hydrodynamic fields from $x=\pm 0$ to
$\pm\infty$, and their discontinuities across the interface, eg
$\Delta T=\delta T_2^B+\delta T_{sq}^B-\delta T_2^A-\delta T_{sq}^A$; cf
Fig.~1. and Eq.~(\ref{sol1a}) for $x=0$. The CoCos of the macroscopic
interface (dotted lines in Fig.~1.) are simpler in three aspects:
First, since it is thicker than the sq-decay length $\lambda_{sq}$, it ends in
a region where the dissipative terms are small and can be neglected. Second,
eliminating dissipative terms especially simplifies $R_s$ and reduces the
number of CoCos, commensurate with the fact that only $\delta T_2^{A,B}$ need
to be determined. Third, the effective discontinuities across the wider
interface include the sq-decay, eg $\Delta_e T=\delta T_2^B-\delta T_2^A$;
cf Eq.~(\ref{sol1a}) for $|x|\gg\lambda_{sq}$. Eqs.~(4) with these
three modifications in-cooperated reduce to \cite{twelve} $\Delta_e f=0$,
$\Delta_e\mu=0$, $\langle f\rangle_e=\kappa_e\Delta_e T$, with an effective
Kapitza conductance $\kappa_e$. They constitute the effective CoCos for
the hydrodynamically wide interface, and are in fact the very CoCos
employed in Ref.~\cite{graliu90} to obtain $\delta T_2^{A,B}$ and
\begin{equation}
 -\rho\dot u=\kappa_e\bigl[\Delta\mu/\langle\sigma\rangle^2\bigr]_i\;.
    \label{res1c}
\end{equation}

The sq-decay was hence implicitly included as a source of
interface dissipation. These previous results therefore remain valid, and the
new information provided by the CoCos, Eqs.~(4), can be seen
by comparing these results with Eqs.~(5), yielding an
expression for $\kappa_e$
\FL
\begin{equation}
 (\kappa_e -{\textstyle\frac{1}{2}}\rho c_2\langle\sigma_T\rangle_i)^{-1}
 =\kappa^{-1}+\sum_{A,B}\bigl(\tilde\alpha\rho c_2\langle\sigma_T\rangle_i
 \bigr)^{-1}\;.    \label{res1b}
\end{equation}

The total effective resistance $\kappa_e^{-1}$ has
four constituting elements: The three in series are on the right hand
side: one microscopic and two sq-contributions. The latter become maximal
for $\alpha_{A,B}=0$, ie if the sq-amplitudes are maximal. The fourth
resistive element, on the left of Eq.~(\ref{res1b}), is circuited in
parallel to the other three. It stems from emission of second sound
which rids the interface of latent heat independent from heat transfer
across the interface. Therefore, this term enables phase
transition even if the actual conductance $\kappa$ vanishes.
(Since its contribution is numerically small, it was not, but should have
been, displayed in Eq.~(5) of Ref.~\cite{graliu90}.) As discussed
above, the actual resistance $1/\kappa$ is
most probably negligible. The experimental data \cite{boyd92} on $\dot u$
then imply $\tilde\alpha_{A,B}\approx 8\cdot 10^2$, if we take
$\alpha_A=\alpha_B$ for lack of better knowledge.

For $\dot u\gg c_2$, the same double approach of actual and effective
CoCos applies. From $R_s$ of Eq.~(\ref{rs}), we obtain (each to the lowest
order of $w/\dot u$ and neglecting cross terms)
\begin{mathletters}\begin{eqnarray}
 g&=&K\Bigl(\langle\sigma\rangle\Delta T+\Delta(\mu_o+z^D)\Bigr)\;,
      \label{cc2a}  \\
 \langle f^D\rangle&=&\beta\Delta T\;,\
                      v_n^A=-\alpha_A(\pi^D-\rho z^D)^A\;,  \label{cc2}
\end{eqnarray}\end{mathletters}

where $\mu_o$ is the chemical potential for a given temperature and
pressure in a system with $v_n=v_s=0$, and $f^D$ the dissipative part
of the entropy current.
The effective CoCos are given by $\Delta_e Q=0$, $\Delta_e\dot\varphi=0$ and
$g=K_e(\langle\sigma\rangle_e\Delta_e T+\Delta_e\mu_o)$.
As partly reported in Ref.~\cite{graliu90}, the latter lead (again via an
expansion around $T_i$) to
\begin{mathletters}\begin{eqnarray}
 \delta T^B&=&-\Bigl[(\Delta\mu_o+T\Delta\sigma)/(T\sigma^B_T)
               \Bigr]_i\;,  \label{res2} \\
 -\rho\dot u&=&K_e\bigl[\Delta\mu_o-{\textstyle\frac{1}{2}}\Delta\sigma
               \delta T^B\bigr]_i\;, \label{res2c} \\
 \delta w^B&=&-(\rho_s/\rho_n\dot u)\bigl[\Delta\mu_o-\sigma^B\delta T^B
               \bigr]_i\;.
\end{eqnarray}\end{mathletters}

(The second equation is valid including $(\delta T^B)^2$.
The proper CoCos for the microscopic interface, Eqs.~(3a,b; 8), provide
the additional information
\FL
\begin{mathletters}\begin{eqnarray}
 \frac{1}{K_e}&=&\left(\frac{1}{K}+\frac{C}{\beta}+\frac{1}{\rho^2\alpha}
                 \right)\left[1+\frac{1}{2}\frac{\Delta\sigma\delta T^B}
                 {\Delta\mu_o}\right]_i\;,  \label{res2a}\\
 \delta w_d^A&=&\frac{\rho_s}{\rho\rho_n}\frac{1}{\alpha}\;\ ,\ \
 \delta T_d^A=\delta T^B\left[1-\frac{\rho\dot u
              \sigma_T^A}{2}\frac{1}{\beta}\right]_i\;,  \label{res2b}
\end{eqnarray}\end{mathletters}

if $\alpha$ and $\beta$ are such that $\delta w_d^A\ll\dot u$,
$\delta T^B-\delta T_d^A\ll\delta T^B$. (Otherwise, the hydrodynamic
dissipation would be too large for the experimental data \cite{boyd92}.)
$C\equiv-{\textstyle\frac{1}{2}}[(\sigma\lambda_w/(\lambda_T-\lambda_w)
-{\textstyle\frac{1}{2}}\Delta\sigma)\sigma_T]_i^A\delta T^B$.
In the first factor of $K_e^{-1}$, three resistive
elements are in series: the first two are microscopic in origin, from
$\Delta\mu$ and $\Delta T$,
respectively; the third is from $w$-diffusion; temperature diffusion
gives rise to the second factor.

\newpage
\vspace*{15mm}
\begin{center}\setlength{\unitlength}{0.240900pt}
\ifx\plotpoint\undefined\newsavebox{\plotpoint}\fi
\sbox{\plotpoint}{\rule[-0.175pt]{0.350pt}{0.350pt}}%
\begin{picture}(1500,809)(0,0)
\sbox{\plotpoint}{\rule[-0.175pt]{0.350pt}{0.350pt}}%
\put(264,158){\rule[-0.175pt]{282.335pt}{0.350pt}}
\put(850,158){\rule[-0.175pt]{0.350pt}{129.604pt}}
\put(1465,158){\makebox(0,0)[l]{$x$}}
\put(850,750){\makebox(0,0){$T(x)-T_i$}}
\put(469,266){\makebox(0,0){$^3He$-$B$}}
\put(1231,266){\makebox(0,0){$^3He$-$A$}}
\put(264,158){\vector(1,0){1172}}
\put(850,158){\vector(0,1){538}}
\put(264,373){\usebox{\plotpoint}}
\put(264,373){\rule[-0.175pt]{109.128pt}{0.350pt}}
\put(717,373){\usebox{\plotpoint}}
\put(717,374){\rule[-0.175pt]{8.431pt}{0.350pt}}
\put(752,374){\usebox{\plotpoint}}
\put(752,375){\rule[-0.175pt]{3.132pt}{0.350pt}}
\put(765,375){\usebox{\plotpoint}}
\put(765,376){\rule[-0.175pt]{2.168pt}{0.350pt}}
\put(774,376){\usebox{\plotpoint}}
\put(774,377){\rule[-0.175pt]{1.445pt}{0.350pt}}
\put(780,377){\usebox{\plotpoint}}
\put(780,378){\rule[-0.175pt]{1.204pt}{0.350pt}}
\put(785,378){\usebox{\plotpoint}}
\put(785,379){\rule[-0.175pt]{0.964pt}{0.350pt}}
\put(789,379){\usebox{\plotpoint}}
\put(789,380){\rule[-0.175pt]{0.723pt}{0.350pt}}
\put(792,380){\usebox{\plotpoint}}
\put(792,381){\rule[-0.175pt]{0.723pt}{0.350pt}}
\put(795,381){\usebox{\plotpoint}}
\put(795,382){\rule[-0.175pt]{0.723pt}{0.350pt}}
\put(798,382){\usebox{\plotpoint}}
\put(798,383){\rule[-0.175pt]{0.482pt}{0.350pt}}
\put(800,383){\usebox{\plotpoint}}
\put(800,384){\rule[-0.175pt]{0.482pt}{0.350pt}}
\put(802,384){\usebox{\plotpoint}}
\put(802,385){\rule[-0.175pt]{0.482pt}{0.350pt}}
\put(804,385){\usebox{\plotpoint}}
\put(804,386){\rule[-0.175pt]{0.482pt}{0.350pt}}
\put(806,386){\usebox{\plotpoint}}
\put(806,387){\rule[-0.175pt]{0.482pt}{0.350pt}}
\put(808,387){\usebox{\plotpoint}}
\put(808,388){\rule[-0.175pt]{0.482pt}{0.350pt}}
\put(810,388){\usebox{\plotpoint}}
\put(810,389){\usebox{\plotpoint}}
\put(811,389){\usebox{\plotpoint}}
\put(811,390){\usebox{\plotpoint}}
\put(812,390){\usebox{\plotpoint}}
\put(812,391){\rule[-0.175pt]{0.482pt}{0.350pt}}
\put(814,391){\usebox{\plotpoint}}
\put(814,392){\usebox{\plotpoint}}
\put(815,392){\usebox{\plotpoint}}
\put(815,393){\usebox{\plotpoint}}
\put(816,393){\usebox{\plotpoint}}
\put(816,394){\usebox{\plotpoint}}
\put(817,394){\usebox{\plotpoint}}
\put(817,395){\usebox{\plotpoint}}
\put(818,395){\usebox{\plotpoint}}
\put(818,396){\usebox{\plotpoint}}
\put(819,396){\usebox{\plotpoint}}
\put(819,397){\usebox{\plotpoint}}
\put(820,397){\usebox{\plotpoint}}
\put(820,398){\usebox{\plotpoint}}
\put(821,398){\usebox{\plotpoint}}
\put(821,399){\usebox{\plotpoint}}
\put(822,399){\usebox{\plotpoint}}
\put(822,400){\usebox{\plotpoint}}
\put(823,400){\usebox{\plotpoint}}
\put(823,401){\usebox{\plotpoint}}
\put(824,401){\usebox{\plotpoint}}
\put(824,402){\usebox{\plotpoint}}
\put(825,402){\usebox{\plotpoint}}
\put(825,403){\usebox{\plotpoint}}
\put(826,404){\usebox{\plotpoint}}
\put(826,405){\usebox{\plotpoint}}
\put(827,405){\usebox{\plotpoint}}
\put(827,406){\usebox{\plotpoint}}
\put(828,406){\rule[-0.175pt]{0.350pt}{0.482pt}}
\put(828,408){\usebox{\plotpoint}}
\put(829,408){\usebox{\plotpoint}}
\put(829,409){\usebox{\plotpoint}}
\put(830,409){\rule[-0.175pt]{0.350pt}{0.482pt}}
\put(830,411){\usebox{\plotpoint}}
\put(831,411){\usebox{\plotpoint}}
\put(831,412){\usebox{\plotpoint}}
\put(832,412){\rule[-0.175pt]{0.350pt}{0.482pt}}
\put(832,414){\usebox{\plotpoint}}
\put(833,414){\rule[-0.175pt]{0.350pt}{0.482pt}}
\put(833,416){\usebox{\plotpoint}}
\put(834,416){\rule[-0.175pt]{0.350pt}{0.482pt}}
\put(834,418){\usebox{\plotpoint}}
\put(835,418){\usebox{\plotpoint}}
\put(835,419){\usebox{\plotpoint}}
\put(836,420){\rule[-0.175pt]{0.350pt}{0.482pt}}
\put(836,422){\usebox{\plotpoint}}
\put(837,422){\usebox{\plotpoint}}
\put(837,423){\usebox{\plotpoint}}
\put(838,424){\rule[-0.175pt]{0.350pt}{0.482pt}}
\put(838,426){\usebox{\plotpoint}}
\put(839,426){\rule[-0.175pt]{0.350pt}{0.482pt}}
\put(839,428){\usebox{\plotpoint}}
\put(840,428){\rule[-0.175pt]{0.350pt}{0.482pt}}
\put(840,430){\usebox{\plotpoint}}
\put(841,431){\rule[-0.175pt]{0.350pt}{0.482pt}}
\put(841,433){\usebox{\plotpoint}}
\put(842,433){\rule[-0.175pt]{0.350pt}{0.723pt}}
\put(842,436){\usebox{\plotpoint}}
\put(843,436){\rule[-0.175pt]{0.350pt}{0.482pt}}
\put(843,438){\usebox{\plotpoint}}
\put(844,439){\rule[-0.175pt]{0.350pt}{0.482pt}}
\put(844,441){\usebox{\plotpoint}}
\put(845,441){\rule[-0.175pt]{0.350pt}{0.723pt}}
\put(845,444){\usebox{\plotpoint}}
\put(846,445){\rule[-0.175pt]{0.350pt}{0.482pt}}
\put(846,447){\usebox{\plotpoint}}
\put(847,447){\rule[-0.175pt]{0.350pt}{0.964pt}}
\put(847,451){\usebox{\plotpoint}}
\put(848,451){\rule[-0.175pt]{0.350pt}{0.723pt}}
\put(848,454){\usebox{\plotpoint}}
\put(849,454){\rule[-0.175pt]{0.350pt}{0.723pt}}
\put(849,457){\usebox{\plotpoint}}
\put(850,458){\usebox{\plotpoint}}
\put(850,158){\rule[-0.175pt]{0.350pt}{72.511pt}}
\put(850,158){\rule[-0.175pt]{141.167pt}{0.350pt}}
\sbox{\plotpoint}{\rule[-0.350pt]{0.700pt}{0.700pt}}%
\put(264,158){\usebox{\plotpoint}}
\put(264,158){\rule[-0.350pt]{141.167pt}{0.700pt}}
\put(850,158){\rule[-0.350pt]{0.700pt}{96.360pt}}
\put(850,558){\usebox{\plotpoint}}
\put(851,558){\rule[-0.350pt]{0.700pt}{0.723pt}}
\put(851,561){\usebox{\plotpoint}}
\put(852,562){\rule[-0.350pt]{0.700pt}{0.723pt}}
\put(852,565){\usebox{\plotpoint}}
\put(853,565){\rule[-0.350pt]{0.700pt}{0.723pt}}
\put(853,568){\usebox{\plotpoint}}
\put(854,568){\rule[-0.350pt]{0.700pt}{0.723pt}}
\put(854,571){\usebox{\plotpoint}}
\put(855,571){\rule[-0.350pt]{0.700pt}{0.723pt}}
\put(855,574){\usebox{\plotpoint}}
\put(856,574){\rule[-0.350pt]{0.700pt}{0.723pt}}
\put(856,577){\usebox{\plotpoint}}
\put(857,577){\rule[-0.350pt]{0.700pt}{0.723pt}}
\put(857,580){\usebox{\plotpoint}}
\put(858,580){\usebox{\plotpoint}}
\put(858,582){\usebox{\plotpoint}}
\put(859,583){\usebox{\plotpoint}}
\put(859,585){\usebox{\plotpoint}}
\put(860,585){\usebox{\plotpoint}}
\put(860,587){\usebox{\plotpoint}}
\put(861,587){\usebox{\plotpoint}}
\put(861,589){\usebox{\plotpoint}}
\put(862,590){\usebox{\plotpoint}}
\put(862,592){\usebox{\plotpoint}}
\put(863,592){\usebox{\plotpoint}}
\put(863,594){\usebox{\plotpoint}}
\put(864,594){\usebox{\plotpoint}}
\put(864,596){\usebox{\plotpoint}}
\put(865,596){\usebox{\plotpoint}}
\put(865,598){\usebox{\plotpoint}}
\put(866,598){\usebox{\plotpoint}}
\put(866,600){\usebox{\plotpoint}}
\put(867,600){\usebox{\plotpoint}}
\put(867,601){\usebox{\plotpoint}}
\put(868,601){\usebox{\plotpoint}}
\put(868,603){\usebox{\plotpoint}}
\put(869,603){\usebox{\plotpoint}}
\put(869,605){\usebox{\plotpoint}}
\put(870,605){\usebox{\plotpoint}}
\put(870,606){\usebox{\plotpoint}}
\put(871,606){\usebox{\plotpoint}}
\put(871,608){\usebox{\plotpoint}}
\put(872,608){\usebox{\plotpoint}}
\put(872,609){\usebox{\plotpoint}}
\put(873,609){\usebox{\plotpoint}}
\put(873,610){\usebox{\plotpoint}}
\put(874,611){\usebox{\plotpoint}}
\put(874,612){\usebox{\plotpoint}}
\put(875,612){\usebox{\plotpoint}}
\put(875,613){\usebox{\plotpoint}}
\put(876,613){\usebox{\plotpoint}}
\put(876,614){\usebox{\plotpoint}}
\put(877,614){\usebox{\plotpoint}}
\put(877,616){\usebox{\plotpoint}}
\put(878,616){\usebox{\plotpoint}}
\put(878,617){\usebox{\plotpoint}}
\put(879,617){\usebox{\plotpoint}}
\put(879,618){\usebox{\plotpoint}}
\put(880,618){\usebox{\plotpoint}}
\put(880,619){\usebox{\plotpoint}}
\put(881,619){\usebox{\plotpoint}}
\put(881,620){\usebox{\plotpoint}}
\put(882,620){\usebox{\plotpoint}}
\put(882,621){\usebox{\plotpoint}}
\put(883,621){\usebox{\plotpoint}}
\put(883,622){\usebox{\plotpoint}}
\put(885,622){\usebox{\plotpoint}}
\put(885,623){\usebox{\plotpoint}}
\put(886,623){\usebox{\plotpoint}}
\put(886,624){\usebox{\plotpoint}}
\put(887,624){\usebox{\plotpoint}}
\put(887,625){\usebox{\plotpoint}}
\put(889,626){\usebox{\plotpoint}}
\put(890,626){\usebox{\plotpoint}}
\put(890,627){\usebox{\plotpoint}}
\put(891,627){\usebox{\plotpoint}}
\put(891,628){\usebox{\plotpoint}}
\put(893,628){\usebox{\plotpoint}}
\put(893,629){\usebox{\plotpoint}}
\put(895,629){\usebox{\plotpoint}}
\put(895,630){\usebox{\plotpoint}}
\put(897,630){\usebox{\plotpoint}}
\put(897,631){\usebox{\plotpoint}}
\put(899,631){\usebox{\plotpoint}}
\put(899,632){\usebox{\plotpoint}}
\put(901,632){\usebox{\plotpoint}}
\put(901,633){\rule[-0.350pt]{0.723pt}{0.700pt}}
\put(904,633){\usebox{\plotpoint}}
\put(904,634){\rule[-0.350pt]{0.723pt}{0.700pt}}
\put(907,634){\usebox{\plotpoint}}
\put(907,635){\rule[-0.350pt]{0.723pt}{0.700pt}}
\put(910,635){\usebox{\plotpoint}}
\put(910,636){\rule[-0.350pt]{0.964pt}{0.700pt}}
\put(914,636){\usebox{\plotpoint}}
\put(914,637){\rule[-0.350pt]{0.964pt}{0.700pt}}
\put(918,637){\usebox{\plotpoint}}
\put(918,638){\rule[-0.350pt]{1.445pt}{0.700pt}}
\put(924,638){\usebox{\plotpoint}}
\put(924,639){\rule[-0.350pt]{1.686pt}{0.700pt}}
\put(931,639){\usebox{\plotpoint}}
\put(931,640){\rule[-0.350pt]{2.650pt}{0.700pt}}
\put(942,640){\usebox{\plotpoint}}
\put(942,641){\rule[-0.350pt]{5.059pt}{0.700pt}}
\put(963,641){\usebox{\plotpoint}}
\put(963,642){\rule[-0.350pt]{113.946pt}{0.700pt}}
\sbox{\plotpoint}{\rule[-0.250pt]{0.500pt}{0.500pt}}%
\put(264,158){\usebox{\plotpoint}}
\put(264,158){\usebox{\plotpoint}}
\put(284,158){\usebox{\plotpoint}}
\put(305,158){\usebox{\plotpoint}}
\put(326,158){\usebox{\plotpoint}}
\put(347,158){\usebox{\plotpoint}}
\put(367,158){\usebox{\plotpoint}}
\put(388,158){\usebox{\plotpoint}}
\put(409,158){\usebox{\plotpoint}}
\put(430,158){\usebox{\plotpoint}}
\put(450,158){\usebox{\plotpoint}}
\put(471,158){\usebox{\plotpoint}}
\put(492,158){\usebox{\plotpoint}}
\put(513,158){\usebox{\plotpoint}}
\put(533,158){\usebox{\plotpoint}}
\put(554,158){\usebox{\plotpoint}}
\put(575,158){\usebox{\plotpoint}}
\put(596,158){\usebox{\plotpoint}}
\put(616,158){\usebox{\plotpoint}}
\put(637,158){\usebox{\plotpoint}}
\put(658,158){\usebox{\plotpoint}}
\put(679,158){\usebox{\plotpoint}}
\put(699,158){\usebox{\plotpoint}}
\put(720,158){\usebox{\plotpoint}}
\put(733,166){\usebox{\plotpoint}}
\put(733,187){\usebox{\plotpoint}}
\put(733,207){\usebox{\plotpoint}}
\put(733,228){\usebox{\plotpoint}}
\put(733,249){\usebox{\plotpoint}}
\put(733,270){\usebox{\plotpoint}}
\put(733,290){\usebox{\plotpoint}}
\put(733,311){\usebox{\plotpoint}}
\put(733,332){\usebox{\plotpoint}}
\put(733,353){\usebox{\plotpoint}}
\put(733,373){\usebox{\plotpoint}}
\put(733,394){\usebox{\plotpoint}}
\put(733,415){\usebox{\plotpoint}}
\put(733,436){\usebox{\plotpoint}}
\put(733,456){\usebox{\plotpoint}}
\put(733,477){\usebox{\plotpoint}}
\put(733,498){\usebox{\plotpoint}}
\put(733,519){\usebox{\plotpoint}}
\put(733,539){\usebox{\plotpoint}}
\put(733,560){\usebox{\plotpoint}}
\put(733,581){\usebox{\plotpoint}}
\put(733,602){\usebox{\plotpoint}}
\put(733,622){\usebox{\plotpoint}}
\put(733,643){\usebox{\plotpoint}}
\put(733,664){\usebox{\plotpoint}}
\put(733,685){\usebox{\plotpoint}}
\put(733,696){\usebox{\plotpoint}}
\put(264,158){\usebox{\plotpoint}}
\put(264,158){\usebox{\plotpoint}}
\put(284,158){\usebox{\plotpoint}}
\put(305,158){\usebox{\plotpoint}}
\put(326,158){\usebox{\plotpoint}}
\put(347,158){\usebox{\plotpoint}}
\put(367,158){\usebox{\plotpoint}}
\put(388,158){\usebox{\plotpoint}}
\put(409,158){\usebox{\plotpoint}}
\put(430,158){\usebox{\plotpoint}}
\put(450,158){\usebox{\plotpoint}}
\put(471,158){\usebox{\plotpoint}}
\put(492,158){\usebox{\plotpoint}}
\put(513,158){\usebox{\plotpoint}}
\put(533,158){\usebox{\plotpoint}}
\put(554,158){\usebox{\plotpoint}}
\put(575,158){\usebox{\plotpoint}}
\put(596,158){\usebox{\plotpoint}}
\put(616,158){\usebox{\plotpoint}}
\put(637,158){\usebox{\plotpoint}}
\put(658,158){\usebox{\plotpoint}}
\put(679,158){\usebox{\plotpoint}}
\put(699,158){\usebox{\plotpoint}}
\put(720,158){\usebox{\plotpoint}}
\put(741,158){\usebox{\plotpoint}}
\put(762,158){\usebox{\plotpoint}}
\put(782,158){\usebox{\plotpoint}}
\put(803,158){\usebox{\plotpoint}}
\put(824,158){\usebox{\plotpoint}}
\put(845,158){\usebox{\plotpoint}}
\put(865,158){\usebox{\plotpoint}}
\put(886,158){\usebox{\plotpoint}}
\put(907,158){\usebox{\plotpoint}}
\put(928,158){\usebox{\plotpoint}}
\put(948,158){\usebox{\plotpoint}}
\put(967,160){\usebox{\plotpoint}}
\put(967,181){\usebox{\plotpoint}}
\put(967,202){\usebox{\plotpoint}}
\put(967,222){\usebox{\plotpoint}}
\put(967,243){\usebox{\plotpoint}}
\put(967,264){\usebox{\plotpoint}}
\put(967,285){\usebox{\plotpoint}}
\put(967,305){\usebox{\plotpoint}}
\put(967,326){\usebox{\plotpoint}}
\put(967,347){\usebox{\plotpoint}}
\put(967,368){\usebox{\plotpoint}}
\put(967,388){\usebox{\plotpoint}}
\put(967,409){\usebox{\plotpoint}}
\put(967,430){\usebox{\plotpoint}}
\put(967,451){\usebox{\plotpoint}}
\put(967,472){\usebox{\plotpoint}}
\put(967,492){\usebox{\plotpoint}}
\put(967,513){\usebox{\plotpoint}}
\put(967,534){\usebox{\plotpoint}}
\put(967,555){\usebox{\plotpoint}}
\put(967,575){\usebox{\plotpoint}}
\put(967,596){\usebox{\plotpoint}}
\put(967,617){\usebox{\plotpoint}}
\put(967,638){\usebox{\plotpoint}}
\put(967,658){\usebox{\plotpoint}}
\put(967,679){\usebox{\plotpoint}}
\put(967,696){\usebox{\plotpoint}}
\end{picture}\end{center}
FIG.~1.\ \ The temperature field for $\dot u\ll c_2$, as in Eq.~(\ref{sol1a}).
\vspace{20mm}
\begin{center}\setlength{\unitlength}{0.240900pt}
\ifx\plotpoint\undefined\newsavebox{\plotpoint}\fi
\sbox{\plotpoint}{\rule[-0.175pt]{0.350pt}{0.350pt}}%
\begin{picture}(1500,809)(0,0)
\sbox{\plotpoint}{\rule[-0.175pt]{0.350pt}{0.350pt}}%
\put(264,158){\rule[-0.175pt]{282.335pt}{0.350pt}}
\put(850,158){\rule[-0.175pt]{0.350pt}{129.604pt}}
\put(1465,158){\makebox(0,0)[l]{$x$}}
\put(850,750){\makebox(0,0){$T(x)-T_i$}}
\put(469,319){\makebox(0,0){$^3He$-$B$}}
\put(1231,319){\makebox(0,0){$^3He$-$A$}}
\put(264,158){\vector(1,0){1172}}
\put(850,158){\vector(0,1){538}}
\put(264,588){\usebox{\plotpoint}}
\put(264,588){\rule[-0.175pt]{141.167pt}{0.350pt}}
\put(850,158){\rule[-0.175pt]{0.350pt}{103.587pt}}
\put(850,158){\rule[-0.175pt]{141.167pt}{0.350pt}}
\sbox{\plotpoint}{\rule[-0.350pt]{0.700pt}{0.700pt}}%
\put(264,158){\usebox{\plotpoint}}
\put(264,158){\rule[-0.350pt]{141.167pt}{0.700pt}}
\put(850,158){\rule[-0.350pt]{0.700pt}{38.785pt}}
\put(850,317){\usebox{\plotpoint}}
\put(850,317){\usebox{\plotpoint}}
\put(851,312){\rule[-0.350pt]{0.700pt}{1.204pt}}
\put(851,312){\usebox{\plotpoint}}
\put(852,307){\rule[-0.350pt]{0.700pt}{0.964pt}}
\put(852,307){\usebox{\plotpoint}}
\put(853,301){\rule[-0.350pt]{0.700pt}{1.204pt}}
\put(853,301){\usebox{\plotpoint}}
\put(854,297){\rule[-0.350pt]{0.700pt}{0.964pt}}
\put(854,297){\usebox{\plotpoint}}
\put(855,292){\rule[-0.350pt]{0.700pt}{0.964pt}}
\put(855,292){\usebox{\plotpoint}}
\put(856,288){\rule[-0.350pt]{0.700pt}{0.723pt}}
\put(856,288){\usebox{\plotpoint}}
\put(857,283){\rule[-0.350pt]{0.700pt}{0.964pt}}
\put(857,283){\usebox{\plotpoint}}
\put(858,279){\rule[-0.350pt]{0.700pt}{0.964pt}}
\put(858,279){\usebox{\plotpoint}}
\put(859,275){\rule[-0.350pt]{0.700pt}{0.723pt}}
\put(859,275){\usebox{\plotpoint}}
\put(860,271){\rule[-0.350pt]{0.700pt}{0.723pt}}
\put(860,271){\usebox{\plotpoint}}
\put(861,267){\rule[-0.350pt]{0.700pt}{0.723pt}}
\put(861,267){\usebox{\plotpoint}}
\put(862,263){\rule[-0.350pt]{0.700pt}{0.964pt}}
\put(862,263){\usebox{\plotpoint}}
\put(863,260){\rule[-0.350pt]{0.700pt}{0.723pt}}
\put(863,260){\usebox{\plotpoint}}
\put(864,256){\rule[-0.350pt]{0.700pt}{0.964pt}}
\put(864,256){\usebox{\plotpoint}}
\put(865,253){\rule[-0.350pt]{0.700pt}{0.723pt}}
\put(865,253){\usebox{\plotpoint}}
\put(866,250){\rule[-0.350pt]{0.700pt}{0.723pt}}
\put(866,250){\usebox{\plotpoint}}
\put(867,247){\rule[-0.350pt]{0.700pt}{0.723pt}}
\put(867,247){\usebox{\plotpoint}}
\put(868,244){\rule[-0.350pt]{0.700pt}{0.723pt}}
\put(868,244){\usebox{\plotpoint}}
\put(869,241){\rule[-0.350pt]{0.700pt}{0.723pt}}
\put(869,241){\usebox{\plotpoint}}
\put(870,238){\rule[-0.350pt]{0.700pt}{0.723pt}}
\put(870,238){\usebox{\plotpoint}}
\put(871,236){\usebox{\plotpoint}}
\put(871,236){\usebox{\plotpoint}}
\put(872,233){\usebox{\plotpoint}}
\put(872,233){\usebox{\plotpoint}}
\put(873,231){\usebox{\plotpoint}}
\put(873,231){\usebox{\plotpoint}}
\put(874,228){\usebox{\plotpoint}}
\put(874,228){\usebox{\plotpoint}}
\put(875,226){\usebox{\plotpoint}}
\put(875,226){\usebox{\plotpoint}}
\put(876,223){\usebox{\plotpoint}}
\put(876,223){\usebox{\plotpoint}}
\put(877,221){\usebox{\plotpoint}}
\put(877,221){\usebox{\plotpoint}}
\put(878,219){\usebox{\plotpoint}}
\put(878,219){\usebox{\plotpoint}}
\put(879,217){\usebox{\plotpoint}}
\put(879,217){\usebox{\plotpoint}}
\put(880,215){\usebox{\plotpoint}}
\put(880,215){\usebox{\plotpoint}}
\put(881,213){\usebox{\plotpoint}}
\put(881,213){\usebox{\plotpoint}}
\put(882,211){\usebox{\plotpoint}}
\put(882,211){\usebox{\plotpoint}}
\put(883,210){\usebox{\plotpoint}}
\put(883,210){\usebox{\plotpoint}}
\put(884,208){\usebox{\plotpoint}}
\put(884,208){\usebox{\plotpoint}}
\put(885,206){\usebox{\plotpoint}}
\put(885,206){\usebox{\plotpoint}}
\put(886,205){\usebox{\plotpoint}}
\put(886,205){\usebox{\plotpoint}}
\put(887,203){\usebox{\plotpoint}}
\put(887,203){\usebox{\plotpoint}}
\put(888,202){\usebox{\plotpoint}}
\put(888,202){\usebox{\plotpoint}}
\put(889,200){\usebox{\plotpoint}}
\put(889,200){\usebox{\plotpoint}}
\put(890,199){\usebox{\plotpoint}}
\put(890,199){\usebox{\plotpoint}}
\put(891,197){\usebox{\plotpoint}}
\put(891,197){\usebox{\plotpoint}}
\put(892,196){\usebox{\plotpoint}}
\put(892,196){\usebox{\plotpoint}}
\put(893,195){\usebox{\plotpoint}}
\put(893,195){\usebox{\plotpoint}}
\put(894,193){\usebox{\plotpoint}}
\put(894,193){\usebox{\plotpoint}}
\put(895,192){\usebox{\plotpoint}}
\put(895,192){\usebox{\plotpoint}}
\put(896,191){\usebox{\plotpoint}}
\put(896,191){\usebox{\plotpoint}}
\put(897,190){\usebox{\plotpoint}}
\put(897,190){\usebox{\plotpoint}}
\put(898,189){\usebox{\plotpoint}}
\put(898,189){\usebox{\plotpoint}}
\put(899,188){\usebox{\plotpoint}}
\put(899,188){\usebox{\plotpoint}}
\put(900,187){\usebox{\plotpoint}}
\put(900,187){\usebox{\plotpoint}}
\put(901,186){\usebox{\plotpoint}}
\put(901,186){\usebox{\plotpoint}}
\put(902,185){\usebox{\plotpoint}}
\put(902,185){\usebox{\plotpoint}}
\put(903,184){\usebox{\plotpoint}}
\put(903,184){\usebox{\plotpoint}}
\put(904,183){\usebox{\plotpoint}}
\put(904,183){\usebox{\plotpoint}}
\put(905,182){\usebox{\plotpoint}}
\put(905,182){\usebox{\plotpoint}}
\put(907,181){\usebox{\plotpoint}}
\put(908,180){\usebox{\plotpoint}}
\put(908,180){\usebox{\plotpoint}}
\put(909,179){\usebox{\plotpoint}}
\put(909,179){\usebox{\plotpoint}}
\put(911,178){\usebox{\plotpoint}}
\put(912,177){\usebox{\plotpoint}}
\put(912,177){\usebox{\plotpoint}}
\put(913,176){\usebox{\plotpoint}}
\put(913,176){\usebox{\plotpoint}}
\put(915,175){\usebox{\plotpoint}}
\put(915,175){\usebox{\plotpoint}}
\put(917,174){\usebox{\plotpoint}}
\put(917,174){\usebox{\plotpoint}}
\put(919,173){\usebox{\plotpoint}}
\put(919,173){\usebox{\plotpoint}}
\put(921,172){\usebox{\plotpoint}}
\put(921,172){\usebox{\plotpoint}}
\put(923,171){\usebox{\plotpoint}}
\put(923,171){\usebox{\plotpoint}}
\put(925,170){\usebox{\plotpoint}}
\put(925,170){\usebox{\plotpoint}}
\put(927,169){\usebox{\plotpoint}}
\put(927,169){\rule[-0.350pt]{0.723pt}{0.700pt}}
\put(930,168){\usebox{\plotpoint}}
\put(930,168){\rule[-0.350pt]{0.723pt}{0.700pt}}
\put(933,167){\usebox{\plotpoint}}
\put(933,167){\rule[-0.350pt]{0.723pt}{0.700pt}}
\put(936,166){\usebox{\plotpoint}}
\put(936,166){\rule[-0.350pt]{0.964pt}{0.700pt}}
\put(940,165){\usebox{\plotpoint}}
\put(940,165){\rule[-0.350pt]{0.964pt}{0.700pt}}
\put(944,164){\usebox{\plotpoint}}
\put(944,164){\rule[-0.350pt]{1.204pt}{0.700pt}}
\put(949,163){\usebox{\plotpoint}}
\put(949,163){\rule[-0.350pt]{1.445pt}{0.700pt}}
\put(955,162){\usebox{\plotpoint}}
\put(955,162){\rule[-0.350pt]{1.686pt}{0.700pt}}
\put(962,161){\usebox{\plotpoint}}
\put(962,161){\rule[-0.350pt]{2.409pt}{0.700pt}}
\put(972,160){\usebox{\plotpoint}}
\put(972,160){\rule[-0.350pt]{3.613pt}{0.700pt}}
\put(987,159){\usebox{\plotpoint}}
\put(987,159){\rule[-0.350pt]{7.709pt}{0.700pt}}
\put(1019,158){\usebox{\plotpoint}}
\put(1019,158){\rule[-0.350pt]{100.455pt}{0.700pt}}
\sbox{\plotpoint}{\rule[-0.250pt]{0.500pt}{0.500pt}}%
\put(264,158){\usebox{\plotpoint}}
\put(264,158){\usebox{\plotpoint}}
\put(284,158){\usebox{\plotpoint}}
\put(305,158){\usebox{\plotpoint}}
\put(326,158){\usebox{\plotpoint}}
\put(347,158){\usebox{\plotpoint}}
\put(367,158){\usebox{\plotpoint}}
\put(388,158){\usebox{\plotpoint}}
\put(409,158){\usebox{\plotpoint}}
\put(430,158){\usebox{\plotpoint}}
\put(450,158){\usebox{\plotpoint}}
\put(471,158){\usebox{\plotpoint}}
\put(492,158){\usebox{\plotpoint}}
\put(513,158){\usebox{\plotpoint}}
\put(533,158){\usebox{\plotpoint}}
\put(554,158){\usebox{\plotpoint}}
\put(575,158){\usebox{\plotpoint}}
\put(596,158){\usebox{\plotpoint}}
\put(616,158){\usebox{\plotpoint}}
\put(637,158){\usebox{\plotpoint}}
\put(658,158){\usebox{\plotpoint}}
\put(679,158){\usebox{\plotpoint}}
\put(699,158){\usebox{\plotpoint}}
\put(720,158){\usebox{\plotpoint}}
\put(733,166){\usebox{\plotpoint}}
\put(733,187){\usebox{\plotpoint}}
\put(733,207){\usebox{\plotpoint}}
\put(733,228){\usebox{\plotpoint}}
\put(733,249){\usebox{\plotpoint}}
\put(733,270){\usebox{\plotpoint}}
\put(733,290){\usebox{\plotpoint}}
\put(733,311){\usebox{\plotpoint}}
\put(733,332){\usebox{\plotpoint}}
\put(733,353){\usebox{\plotpoint}}
\put(733,373){\usebox{\plotpoint}}
\put(733,394){\usebox{\plotpoint}}
\put(733,415){\usebox{\plotpoint}}
\put(733,436){\usebox{\plotpoint}}
\put(733,456){\usebox{\plotpoint}}
\put(733,477){\usebox{\plotpoint}}
\put(733,498){\usebox{\plotpoint}}
\put(733,519){\usebox{\plotpoint}}
\put(733,539){\usebox{\plotpoint}}
\put(733,560){\usebox{\plotpoint}}
\put(733,581){\usebox{\plotpoint}}
\put(733,602){\usebox{\plotpoint}}
\put(733,622){\usebox{\plotpoint}}
\put(733,643){\usebox{\plotpoint}}
\put(733,664){\usebox{\plotpoint}}
\put(733,685){\usebox{\plotpoint}}
\put(733,696){\usebox{\plotpoint}}
\put(264,158){\usebox{\plotpoint}}
\put(264,158){\usebox{\plotpoint}}
\put(284,158){\usebox{\plotpoint}}
\put(305,158){\usebox{\plotpoint}}
\put(326,158){\usebox{\plotpoint}}
\put(347,158){\usebox{\plotpoint}}
\put(367,158){\usebox{\plotpoint}}
\put(388,158){\usebox{\plotpoint}}
\put(409,158){\usebox{\plotpoint}}
\put(430,158){\usebox{\plotpoint}}
\put(450,158){\usebox{\plotpoint}}
\put(471,158){\usebox{\plotpoint}}
\put(492,158){\usebox{\plotpoint}}
\put(513,158){\usebox{\plotpoint}}
\put(533,158){\usebox{\plotpoint}}
\put(554,158){\usebox{\plotpoint}}
\put(575,158){\usebox{\plotpoint}}
\put(596,158){\usebox{\plotpoint}}
\put(616,158){\usebox{\plotpoint}}
\put(637,158){\usebox{\plotpoint}}
\put(658,158){\usebox{\plotpoint}}
\put(679,158){\usebox{\plotpoint}}
\put(699,158){\usebox{\plotpoint}}
\put(720,158){\usebox{\plotpoint}}
\put(741,158){\usebox{\plotpoint}}
\put(762,158){\usebox{\plotpoint}}
\put(782,158){\usebox{\plotpoint}}
\put(803,158){\usebox{\plotpoint}}
\put(824,158){\usebox{\plotpoint}}
\put(845,158){\usebox{\plotpoint}}
\put(865,158){\usebox{\plotpoint}}
\put(886,158){\usebox{\plotpoint}}
\put(907,158){\usebox{\plotpoint}}
\put(928,158){\usebox{\plotpoint}}
\put(948,158){\usebox{\plotpoint}}
\put(967,160){\usebox{\plotpoint}}
\put(967,181){\usebox{\plotpoint}}
\put(967,202){\usebox{\plotpoint}}
\put(967,222){\usebox{\plotpoint}}
\put(967,243){\usebox{\plotpoint}}
\put(967,264){\usebox{\plotpoint}}
\put(967,285){\usebox{\plotpoint}}
\put(967,305){\usebox{\plotpoint}}
\put(967,326){\usebox{\plotpoint}}
\put(967,347){\usebox{\plotpoint}}
\put(967,368){\usebox{\plotpoint}}
\put(967,388){\usebox{\plotpoint}}
\put(967,409){\usebox{\plotpoint}}
\put(967,430){\usebox{\plotpoint}}
\put(967,451){\usebox{\plotpoint}}
\put(967,472){\usebox{\plotpoint}}
\put(967,492){\usebox{\plotpoint}}
\put(967,513){\usebox{\plotpoint}}
\put(967,534){\usebox{\plotpoint}}
\put(967,555){\usebox{\plotpoint}}
\put(967,575){\usebox{\plotpoint}}
\put(967,596){\usebox{\plotpoint}}
\put(967,617){\usebox{\plotpoint}}
\put(967,638){\usebox{\plotpoint}}
\put(967,658){\usebox{\plotpoint}}
\put(967,679){\usebox{\plotpoint}}
\put(967,696){\usebox{\plotpoint}}
\end{picture}\end{center}
FIG.~2.\ \ The temperature field for $\dot u\gg c_2$, as in Eq.~(\ref{sol2a}).

\begin{references}
\bibitem{nine} This is a ballistic result that does not extend into the
 collision-dominated hydrodynamic regime. On hydrodynamic length scales,
 any $v_n$ that traverses the interface induces $\partial_x v_n$,
 and hence a temperature variation, via $\alpha_{A,B}$ of Eq.~(\ref{cc1}).
 This large scale dissipation will always feed back to the interface motion,
 even if it starts at $\Delta\mu=\Delta T=0$, and halt it eventually.
\bibitem{twelve} Continuity of energy flux, $\Delta Q=0$, and phase coherence,
 $\Delta\dot\varphi=0$, are quite generally valid stationarily irrespective of
 the interface width. Also cf Ref.~\cite{paliu92}.
\end{references}
\end{document}